\documentclass{appolb}
\usepackage{epsfig}

\begin{document}
\title{The First 3 Years at RHIC- an Overview%
\thanks{Presented at the XXXIV International Symposium on 
Multiparticle Dynamics, Sonoma State University, California, USA}%
}
\author{Richard Seto
\address{University of California, Riverside}
}
\maketitle
\begin{abstract}
This is an overview of the results from the first 3 years of RHIC
experiments.  RHIC is a collider built to accelerate nuclei to center
of mass energies of 200 GeV per nucleon for the study of QCD in bulk
systems. The most important result so far is the observation of the
suppression of high p$_T$ hadrons in central Au-Au collisions followed
by the subsequent null experiment where the same suppression was not
seen in deuteron-Au collisions. The observed suppression is a
final state effect in which a large amount of energy is lost by the
fast parton as it penetrates the medium. This observation, together
with measurements of the elliptic flow, leads to the conclusion that
the energy density reached is at least 10 times that of a normal
nucleon. The simplest and most economical explanation of these
phenomenon is that the system is a dense, locally thermalized system
of unscreened color charges.
\end{abstract}
  
\section{Introduction}
QCD (Quantum Chromodynamics) is the established theory of the strong
interactions which together with the electro-weak force constitute the
forces in the Standard Model. These, and perhaps all fundamental
theories picture the vacuum as a complex sea of stuff. It is the
interaction of the fundamental constituents of the theories with the
vacuum that generates mass and, in the case of the strong interaction
(QCD) gives rise to the phenomenon of quark confinement.  Since the
vacuum is a medium, its structure can be altered as the temperature is
changed\cite{Lee}. The most violent of these changes is a phase
transition. The various components of the vacuum corresponding to the
various forces each went through one or more such transitions at their
characteristic temperatures early in the history of the universe. It
is very likely that such a phase transition powered the sudden
expansion of the universe known as inflation. Most of the phase
transitions studied experimentally - e.g.  water to ice, Helium-3,
magnetic domains - all result from the electromagnetic force. It is
natural to ask if we can study a phase transition or transitions
resulting from QCD. Such was the task of the Relativistic Heavy Ion
Collider (RHIC) which began taking data in 2001. This talk is an
overview of what we have learned from the first three major data
taking periods. RHIC, located at the Brookhaven National Laboratory,
is a large accelerator which can accelerate heavy ions of all species
to center of mass energies of 200 GeV.  Such violent collisions of
large nuclei will leave in its wake, a region of high energy density
whose net baryon-density is nearly zero - that is, a high temperature
vacuum.

In the past year, the RHIC community, with its four major experiments,
STAR, PHENIX, PHOBOS and BRAHMS,  has been taking stock of the of
its status, in regards to the the discoveries made, and the questions
yet to be answered\cite{white}. The task was organized around 4 questions:
\begin{enumerate}
\item Does the system formed at RHIC reach thermal and chemical
equilibrium? If so, what is the initial temperature or energy density?
How does the system evolve?
\item Have we seen the signatures of the deconfinement phase transition? If
so, what have we learned about the mechanism of confinement?
\item Have we observed the chiral phase transition? What is the relationship
between the QCD vacuum and the masses of the hadrons? Or in other
words - what is the origin of chiral symmetry breaking?
\item What are the properties of matter at very high energy densities? Is
the quark and gluon description the best way to understand the system?
\end{enumerate}

Before describing the experimental data, it is worth reviewing some of
the expectations and ideas from theory. Quantitative predictions of
QCD at momentum transfers below 1 GeV are difficult, since the
coupling constant is large and perturbation theory will not
work. Lattice gauge calculations, often done on powerful computers,
are used to obtain numerical values for quantities of interest such as
the hadron masses spectrum\cite{lattice-mass}. Such calculations can
also be used to predict the critical temperature for the phase
transition. QCD has at least two transitions, which are in all
likelihood connected. The first is the deconfinement transition, in
which the quarks are set free from the confines of their parent
hadrons. The second is the chiral transition - the transition
responsible for the bulk of the hadronic mass\footnote{The masses of
the quarks, at temperatures above the chiral transition are ~ several
MeV and can be taken to be nearly zero. In this case, the left and
right handed sector of quarks are completely separate - hence the name
``chiral'' symmetry. At low temperature, quarks attain a ``dressed''
mass due to their interaction with the vacuum and chiral symmetry is
broken.}.  The two transitions are thought to be at the same
temperature. Lattice calculations predict a critical temperature T$_C$
of about 170 MeV, giving a critical energy density $\epsilon_C\sim
1~GeV/fm^3$
\cite{lattice-temp}.  In a theory with only gluons and no quarks, the
transition is first order. In nature, since the u and d quarks have a
small mass, and the strange quark has a somewhat larger mass, the
phase transition is predicted to be a cross over. However, since this
cross over occurs over a very narrow range of temperatures, the
transition is, for all practical purposes, first order, since the
temperature cannot be controlled to anywhere near the accuracy needed
to tell the difference.

\section{Preliminaries}
The ideal experiment would be to make a trap for nuclear matter and
raise the temperature, as in the center of a star. Unfortunately,
there is no containment mechanism which can withstand such forces in
the laboratory. Such enormous pressures and temperatures can be
produced in high energy collisions from an accelerator. However the
duration that the relevant state exists is very short, and
evolves with time.  Experimentalists must examine the debris of such a
collision, whose products come from every stage of the system - both
above the transition temperature and below. Some of the products will
come from the initial collision before equilibration is reached, and
other products will come from reactions taking place significantly
after the system has cooled below the transition temperature and will
give a background to the interesting products made during the high
temperature phase.  One of the key ideas is to utilize experimental
probes which give information about particular time periods in the
evolution of the system. The RHIC experiments become an archaeological
expedition, albeit, the timescale is rather short - less then
$10^{-21}$ seconds. But like the archaeologist we must be able to date
the relics that we find.

\begin{figure}
\begin{minipage}[t]{0.54\linewidth}
\centering
\includegraphics[width=2.5in]{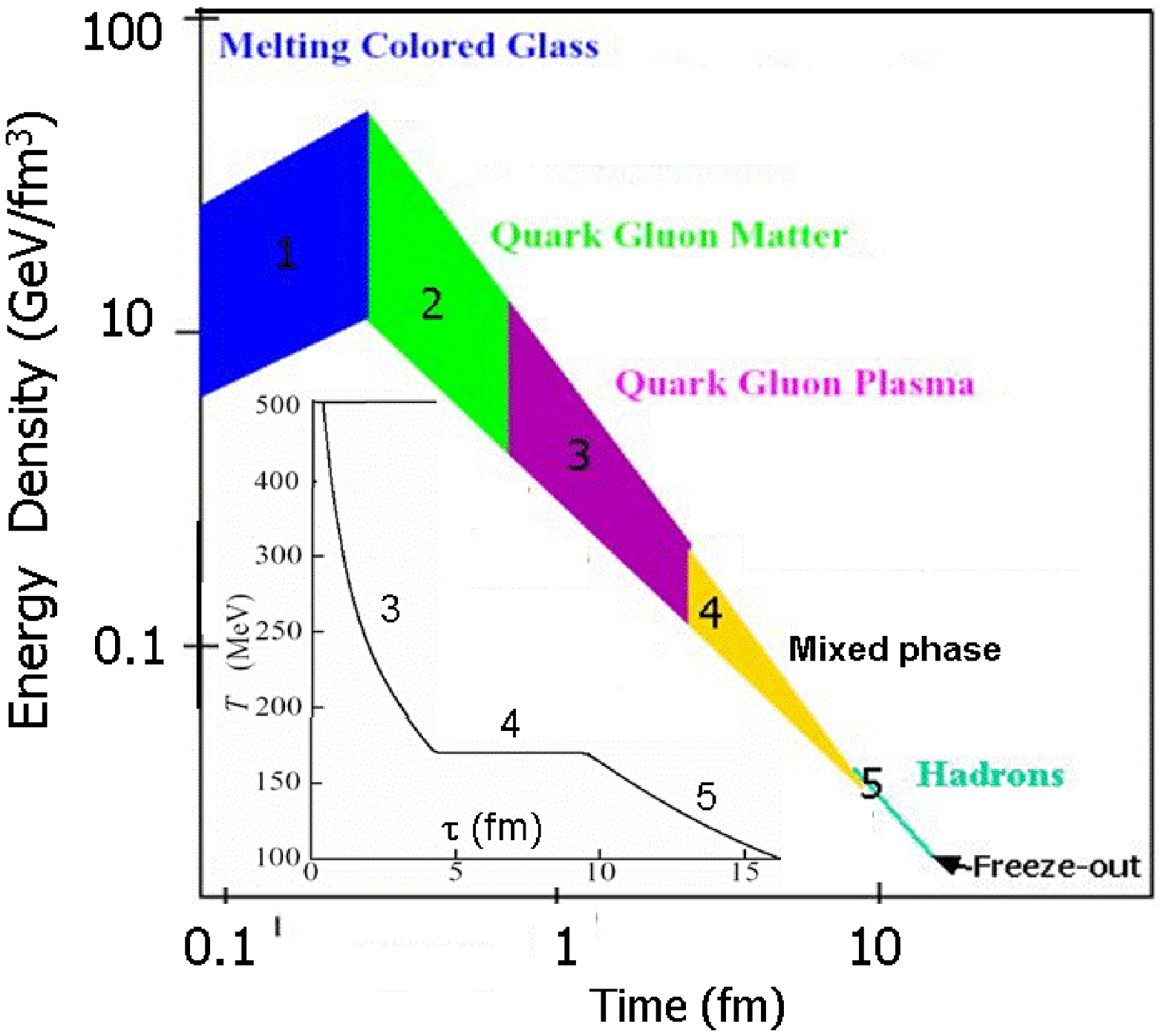}
\caption{A schematic of the energy density vs time history of the systems 
studied at RHIC in Au-Au Collisions. The five stages are described in
the text. The inset is a temperature vs time history for the stages
where the temperature is reasonably well defined. Although the figure shows
what appears to be a first order transition, the best
estimates from lattice calculations tell us that with 2 light quarks,
and a moderately heavy strange quark, the transition is a cross over
which occurs over a relatively narrow band of temperatures. (Figure
due to Larry McLerran.)} 
\label{fig:larry}
\end{minipage}%
\hspace{.2in}%
\begin{minipage}[t]{0.44\linewidth}
\centering
\includegraphics[width=2.1in]{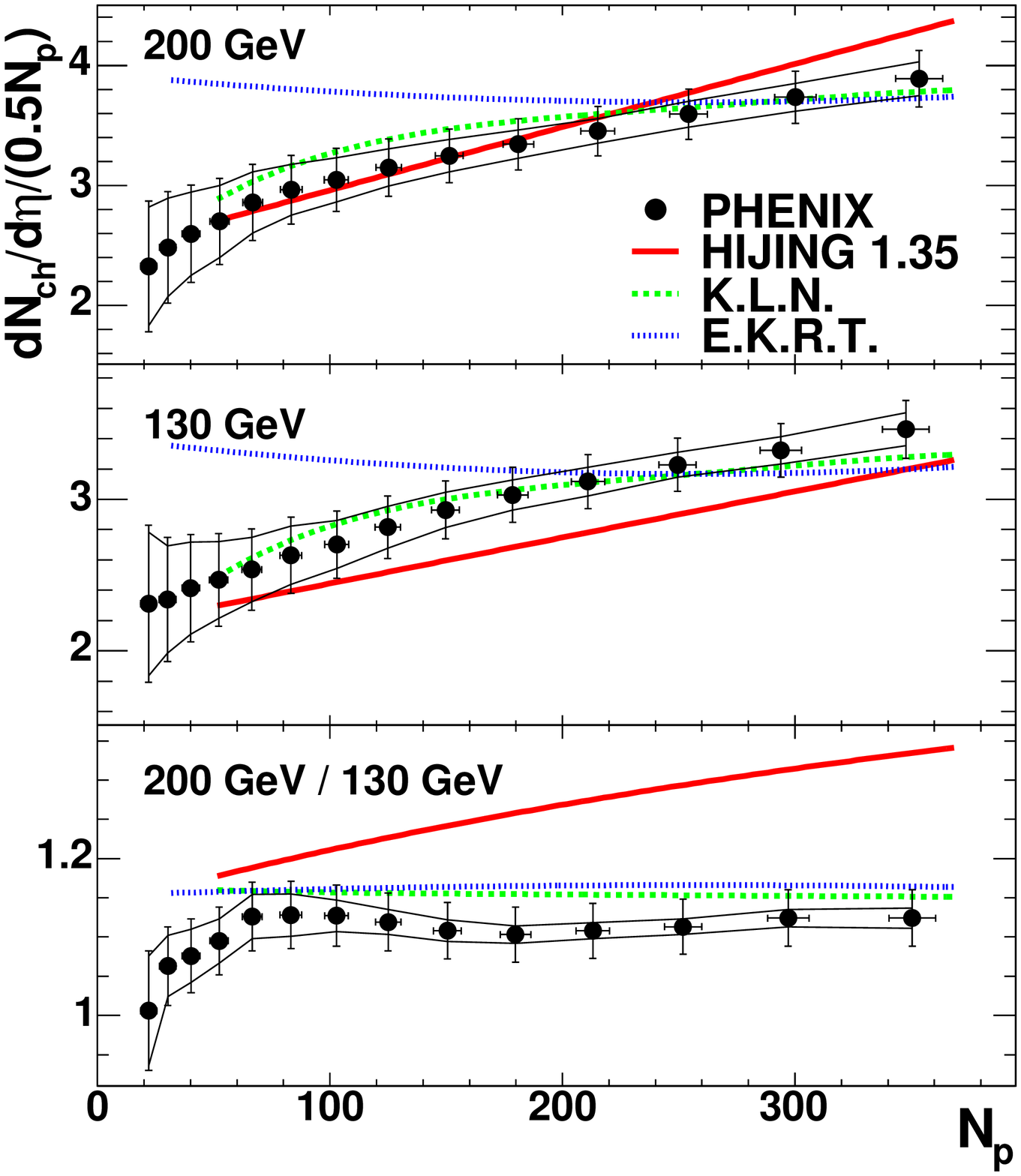}
\caption{The charged particle multiplicity per unity pseudorapidity, 
normalized to N$_{part}$ at 200 GeV and 130 GeV center of mass energy
for Au-Au collisions as a function of N$_{part}$. K.L.N. is the model
by Kharzeev, Levin, and Nardi\cite{KLN} based on saturation
ideas. E.K.R.T. refers to a model which assumes saturation in the
final state\cite{EKRT}, and Hijing 1.35\cite{hijing} is a model based
on pQCD mini-jets. }
\label{fig:dndeta}
\end{minipage}
\end{figure}

Fig. \ref{fig:larry} is a cartoon of the evolution of a heavy ion
collision showing the energy density vs time on a log scale. The
inset shows the temperature as a function of the time. If one assumes
a Stefan-Boltzmann relationship between the energy density and the
temperature then one has $\epsilon_{SB}=N_{DOF}\frac{\pi^2}{30}T^4$
where the $N_{DOF}$ is the number of degrees of freedom in the system
which ranges from 37 to about 47.5 depending on whether the strange
quark is taken as massless.  Unlike traditional particle physics
experiments - we are not interested in processes involving only a
single scattering. Rather we are interested in the many body processes
of the bulk where concepts such as a local temperature and entropy
have meaning. The collision proceeds in 5 stages.
\begin{enumerate}
\item The initial state - a Colored Glass Condensate - so named because 
of a model in which the color fields are studied in a classical
approximation because of the high occupation numbers. The fields of
fast moving particles serve as sources for the slower fields and
provide the ``frustration'' that is typical of a glass. The large
Q$^2$ processes occurring during this stage provide the high
p$_T$ probes of the system useful for studying later stages of the
collision.
\item Quark gluon matter - a pre-equilibrium stage of quarks and 
gluons lasting about 1 fm leading to a locally equilibrated system.
The phenomenon of elliptic flow begins to develop at the end of this stage.
\item The Quark Gluon Plasma phase. Hard probes leading to the phenomenon 
of jet suppression are particularly important here, since most
of the their energy is lost in this phase.
\item The mixed phase, which presumably spans a narrow range in temperature 
and hadrons form. Even if the transition is strictly a cross over, it
is assumed that there is some time during the collision where there is
a mixture of quark-gluon-plasma and hadrons, while the hadrons are
forming. 
It is also in this phase, together with the QGP phase where chiral
symmetry is restored.
\item The hadronic phase where hadrons interact with one another. Radial
 flow develops and ends in freezeout of the final hadronic products
 (meaning that all interactions between particles cease) setting the
 kinetic and chemical freezeout temperatures.
\end{enumerate}

We can divide the processes in a relativistic heavy ion collision into
two categories - hard and soft. Hard processes are those with a
p$_T>3$ GeV, where perturbative calculations are reasonably
accurate. Hard processes can be used as a calibrated probe of the
medium since they should scale as the number of initial parton-parton
collisions. Shadowing will give a correction to this which must be
accounted for by studying proton (or deuteron) nucleus collisions. The
modeling of hard processes follows the standard methods of pQCD
calculations using structure functions followed by jet
fragmentation. These hard processes provide high momentum
partons, which loose energy in the medium. This phenomenon, known
as jet quenching - is one of the major experimental signatures seen at
RHIC.

Models of the soft processes are more complicated since they are
non-perturbative. These are the processes which lead to the majority of
particle production - and hence to the quark-gluon plasma. Non-viscous
hydrodynamics, which assumes that the bulk matter is a continuous
medium, is often used to model the evolution of the
system. Hadronization is done using the so called ``Cooper Frye''
formalism which simply converts the continuous matter to hadrons
conserving charge, momentum and energy in a Lorentz invariant
manner. Hydrodynamics requires two external inputs- the initial
conditions, and the equation of state.  For the latter of these, one
can simply assume the EOS of an ideal gas -either in the hadronic
stage, in which the degrees of freedom are the hadrons, or the QGP
stage in which the degrees of freedom are the quarks and gluons.
These are often taken as limiting cases and a variety of EOS's are
tested. The assumption of zero-viscosity will turn out to be
important, as this implies that the medium is actually not an ideal
gas, but is rather strongly interacting. In hindsight - that this is
true might seem to be obvious as the value of $\alpha_S$ at
$\sqrt{Q^2}\sim T \sim 300$ MeV is rather large during the QGP phase.

Recently, Mclerran\cite{larry} and his collaborators have used a
classical approximation for the initial stage of the collision,
arguing that the occupation numbers at low x where much of the
particle production occurs are rather high. This model - which they
have named the ``Colored Glass Condensate''- shows the phenomenon of
gluon saturation and makes predictions which can be used to calculate
the initial conditions in a heavy ion collision which in turn can then
be used as input to the hydrodynamical calculations. This calculation
relies on the fact that very early in the collision, gluon saturation
effects at low x set a value of Q$\sim$Q$_S$ where $\alpha_S$ can be considered
small but the occupation numbers are high. The value of $Q_S$ at RHIC
is 1-2 GeV so $\alpha_S^2\sim\frac{1}{10}$. The
saturation assumed by these authors is present in the initial
state before the nuclei collide. This fact will be important in
distinguishing these effects, from final state effects such as the
formation of a quark-gluon plasma.

A second recent advance has to do with the later stages of the
collision - hadronization. Several groups have conjectured
that low momentum hadrons in the final state, come primarily from the
recombination of partons and not from the fragmentation. While a
rigorous calculation can only be done in a range of momenta
where masses can be ignored, the general trends predicted by these
models seem to explain a variety of experimental observables between
about 1 and 4 GeV/c momentum\cite{recomb}.

\begin{figure}
\begin{center}
\includegraphics[width=2.3in, angle=90]{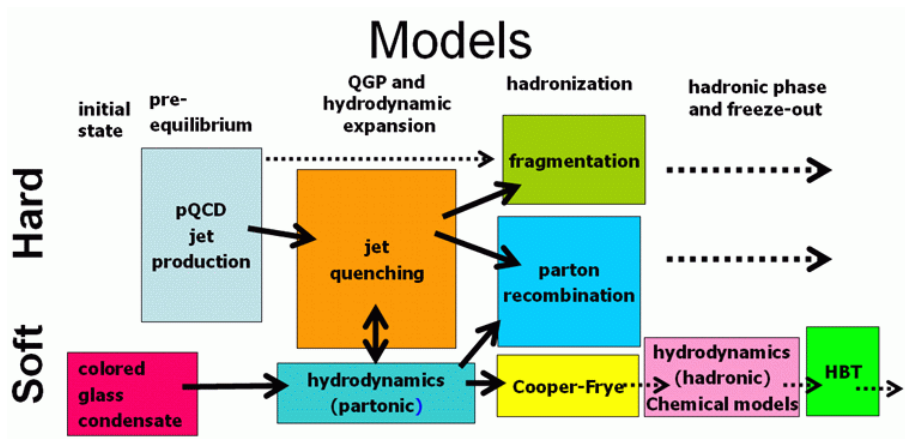}
\caption{A diagram of models used for understanding 
collisions at RHIC. Different approximations are appropriate in
each regime depending on the scale of the interaction being
modeled, and the degree of thermalization (figure due to S. Bass).}
\label{fig:model}
\end{center}
\end{figure}

Fig.~\ref{fig:model} shows a schematic of these various models and
their connections. The division in the applicability
of these models is not absolute. In fact, hydrodynamics which provides
the framework for understanding the motion of low and moderate
momentum partons will feed into parton recombination models. Parton
recombination, which was initially believed to recombine only thermal
partons, appear also to incorporate low momentum partons from jet
fragmentation as well. For a complete understanding of the
experimental data, a rather sophisticated picture, involving all of
these models is necessary. This processes is still in its infancy and
the cooperation of various types of theorists and
experimentalists will be needed to gain a detailed understanding of
the dynamics of the heavy ion collision. In the initial stages of this
task - it is important that we concentrate on whether the
overall ideas are correct - even if all the experimental data is not
fully reproduced by the models.

One of the important control parameters used by the RHIC experiments,
is the impact parameter or centrality of the collision. By convention
0\% centrality refers those collisions having the smallest impact
parameter, and the term peripheral refers to glancing collisions.  The
four RHIC experiments all have identical devices (the zero degree
calorimeters) which measure the centrality. Soft processes generally
scale with the number of participating nucleons in the collisions or
N$_{part}$ which is a gross measure of the size and/or energy density
of the fireball, whereas the number of hard interactions scale with
the number of collisions or N$_{coll}$. Once the impact parameter is
determined via measurements of the zero-degree calorimeters, a simple
Glauber model is used to determine N$_{part}$ and N$_{coll}$.

\section{The Initial State - a Colored Glass Condensate}
One of the surprising (and for some, disturbing) early observations,
was that the multiplicities coming from heavy ion collisions at RHIC
energies, was lower than many of the predictions coming from naive
pQCD estimates. Kharzeev and his colleagues used the Colored Glass
Condensate model to make a prediction of the multiplicity as a
function of centrality. They obtained
$\frac{dN}{dy}\sim\frac{1}{\alpha(Q_S)}N_{part}$ where Q$_S$ is the
saturation momentum which is a slow function of N$_{part}$ coming from
the fact that the particle density and hence the saturation scale is
dependent on the centrality. Fig.~\ref{fig:dndeta} shows a comparison
of three models with multiplicity data from PHENIX. One can see that
the model based on saturation, by Kharzeev, Levin and Nardi, (labeled
K.L.N.)\cite{KLN} makes a reasonable accounting for the data at both
200 and 130 GeV center of mass. The paucity of particles compared to
naive expectations is attributed to saturation which limits particle
production. What is somewhat disconcerting is that the calculation
seems to work reasonably at a $\sqrt{s}$=19.6 GeV, where one might not
expect the model to be valid.  Whether this is cause to doubt the
model remains to be seen.  In any case one can extract from these
models, an energy density in the early stages of the collision of
about 18 GeV/fm$^3$\cite{KLN} well above the lattice value of 1
GeV/fm$^3$ required for the phase transition.

\begin{figure}
\begin{center}
\includegraphics[width=4.5 in]{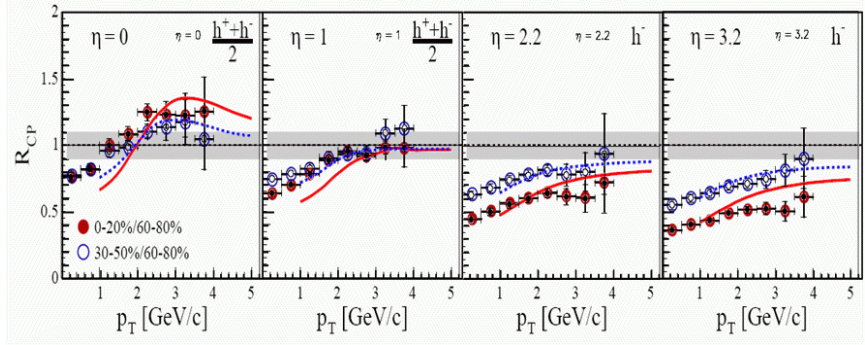}
\caption{The ratio R$_{CP}$ for charged hadrons in deuteron-Au collisions for
central events and mid-central events in varying bins of
pseudorapidity, showing the Cronin enhancement at mid-rapidity and
suppression at forward rapidity (Brahms). The fits are to a model by Tuchin et
al.\cite{tuchin}}
\label{fig:brahms}
\end{center}
\end{figure}

A second piece of evidence for the CGC relies on the fact that in
going to forward rapidities, one begins to sample a lower range in x
in the nucleus.  The ratio
$R_{CP}=\frac{Yield(central)/N_{coll}(Central)}{Yield(peripheral)/N_{coll}(Peripheral)}$
is a measure of the yield per collision from hard processes coming
from central as compared to peripheral collisions, where the
peripheral collisions are taken as a baseline. If pp data is
available, it is often used as the baseline as will be done later in
the definition of R$_{AA}$. The BRAHMS experiment, whose strength is
the capability to measure very forward rapidities, looked at this
ratio in deuteron-nucleus collisions. For a given $p_T$, a lower and
lower value of x is sampled as one moves to higher rapidity. Since the
gluon structure function increases at low x one would see a stronger
suppression as one moves to higher rapidity. Fig.~\ref{fig:brahms}
shows just this effect, with the more central collisions showing a
larger suppression as one might expect.  At midrapidity, above a p$_T$
of 2 GeV, one sees an enhancement instead of a suppression. This
phenomenon, known as the Cronin effect, comes from initial state
multiple scattering of the incoming projectile parton. Even with this
enhancement, the saturation effects are strong enough to show an
overall suppression at forward rapidities of a factor of
2. Theoretical saturation calculations by Kharzeev, Kovchegov and
Tuchin show a similar qualitative trend.

\section{Thermalization and Elliptic Flow}
One of the surprising results which was immediately apparent at RHIC
was a strong directional anisotropy in momentum known as elliptic
flow. Initially the concept may seem foreign to particle physicists,
but this is akin to other measurements that have been used in the
study of the strong interactions - the ``jet shape'' variables of
thrust, and sphericity. In the case of heavy ion collisions, such
behavior involves all particles emerging from the
interaction and has nothing to do with jets or hard scattering, but
arises from pressure gradients in a spatially anisotropic collision.
The anisotropy is strongest in mid-central collisions and disappears
for very peripheral or very central collisions.  The conversion
efficiency from spacial to momentum anisotropy depends on the
properties of the medium and hence can be use to understand its
properties.  In order for efficient conversion, the medium must be
strongly coupled.  Contrary to what one might presuppose, this implies
a zero viscosity and zero mean-free path. Such systems, often called
perfect fluids, have been studied in other areas such as atomic
physics\cite{science}. One can also calculate the viscosity in
particular strongly coupled theories using the AdS5/CFT duality- where
one finds that the viscosity ~zero\cite{adscft}.  In relativistic
heavy ion physics, non-dissipative hydrodynamics is used. The quantity
of interest is the value of the second Fourier coefficient of the
azimuthal momentum anisotropy - the elliptic flow. In simple terms it
is the extent to which the shape is elliptical as opposed to
spherical.  One of the important external inputs to these models is
the initial thermalization time at which the pressure gradients begin
to be operational. Before this time, the system is assumed to free
stream and expand isotropically reducing the spacial anisotropy and
thereby the elliptic flow. Using this fact, the value of the elliptic
flow when compared to the spacial anisotropy which one obtains from
centrality measurements, can give an estimate of the thermalization
time. It is found that thermalization times of about 0.6 to 1 fm are
required to fit the data. In these models one obtains an energy
density of 15-25 GeV/fm$^3$\cite{v$_2$energydensity} similar to the
estimate given from the CGC initial conditions.


\section{Jet quenching}
\begin{figure}
\begin{center}
\includegraphics[width=2.7in, angle=-90]{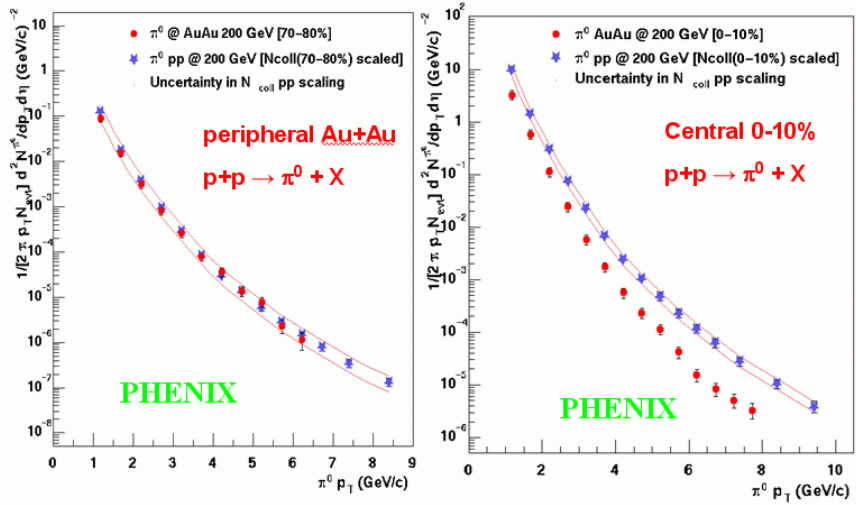}
\caption{Left: The p$_T$ spectrum of $\pi^0$'s normalized to N$_{coll}$ 
for pp collisions compared to peripheral AuAu collisions. Right: The same
for central AuAu collisions, showing that the Au-Au data do not scale -
the effect of parton energy loss\cite{quench}.}
\label{fig:jetquench}
\end{center}
\end{figure}

A long sought signal of high density matter has been the large loss of
energy of a fast parton as it penetrates the medium. The energy loss can
easily be understood as the radiation of gluons from the fast parton, because
of the strong color charges.  Since
high p$_T$ particles from hard processes scale as $N_{coll}$,
one simply compares the p$_T$ spectrum measured in central heavy ion
collisions scaled by N$_{coll}$ to a baseline measured in pp
collisions.  This effect was dramatically seen at RHIC as shown in
Fig.~\ref{fig:jetquench} where one can see the rather large (factor of
~4-5) suppression for high p$_T$ $\pi^0$'s as compared to the pp
scaled expectation. One can also see that for peripheral
collisions, the scaling works rather well. The N$_{coll}$ scaling of
hard processes has been double checked using direct photons, which are
produced via hard processes but do not loose energy in the medium
since they have no color charge. \footnote{It also appears that
single electrons coming primarily from charm 
(after the dalitz and photon-conversion contributions
have been subtracted) follow this
scaling as well. Heavy quarks are thought not to loose energy due to a
dead cone effect that limits the radiation because of kinematics.}

\begin{figure}
\begin{minipage}[t]{0.48\linewidth}
\centering
\includegraphics[width=2.4in]{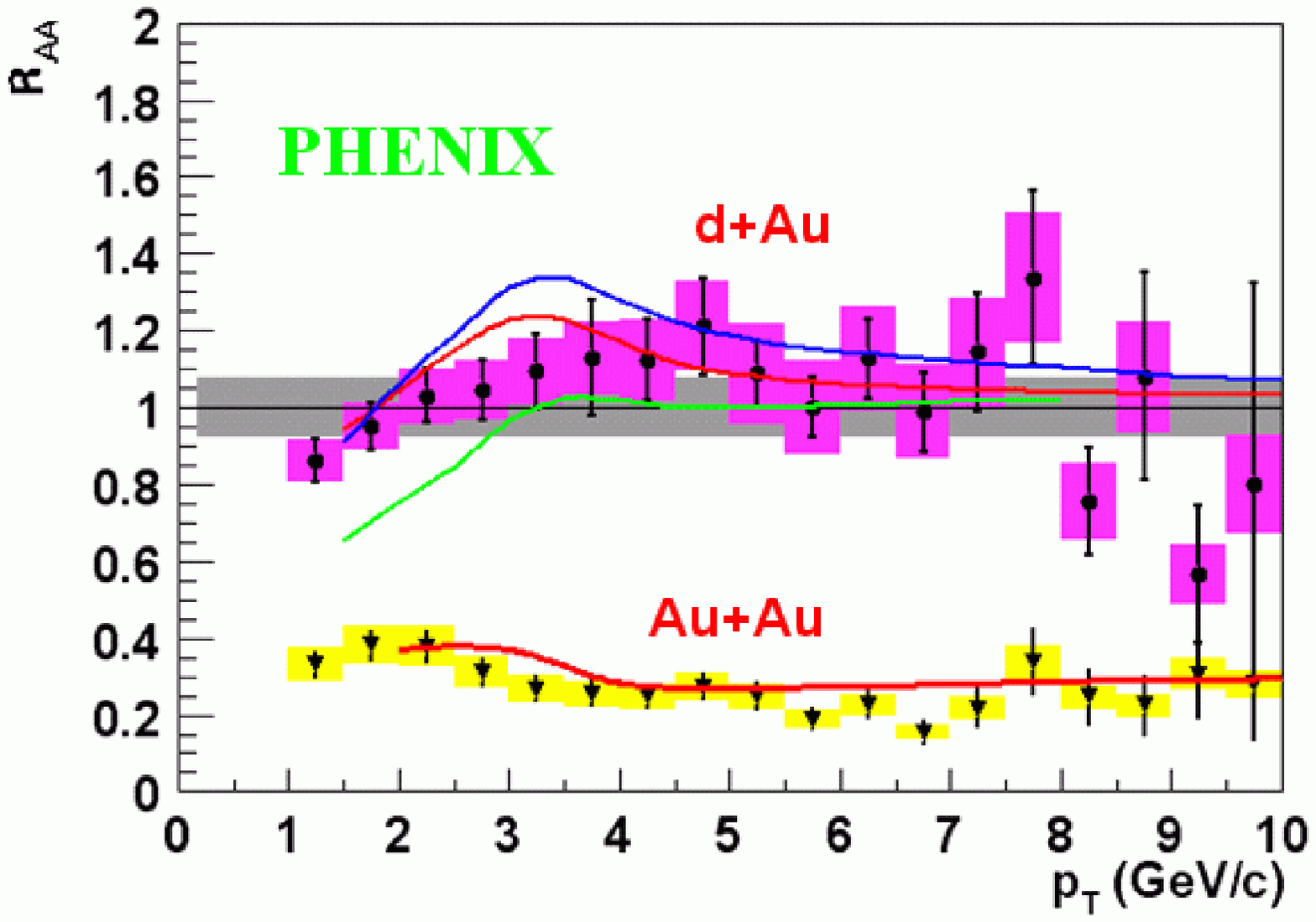}
\caption{The ratio R$_{AA}$ as explained in the text for dAu collisions
and central AuAu collisions. One can see the clear suppression below
unity for central AuAu data, and a lack of suppression in the dAu
data.\cite{quench,quench_da,quench_models}} \label{fig:raa}
\end{minipage}%
\hspace{.2in}%
\begin{minipage}[t]{0.48\linewidth}
\centering
\includegraphics[width=2.4in]{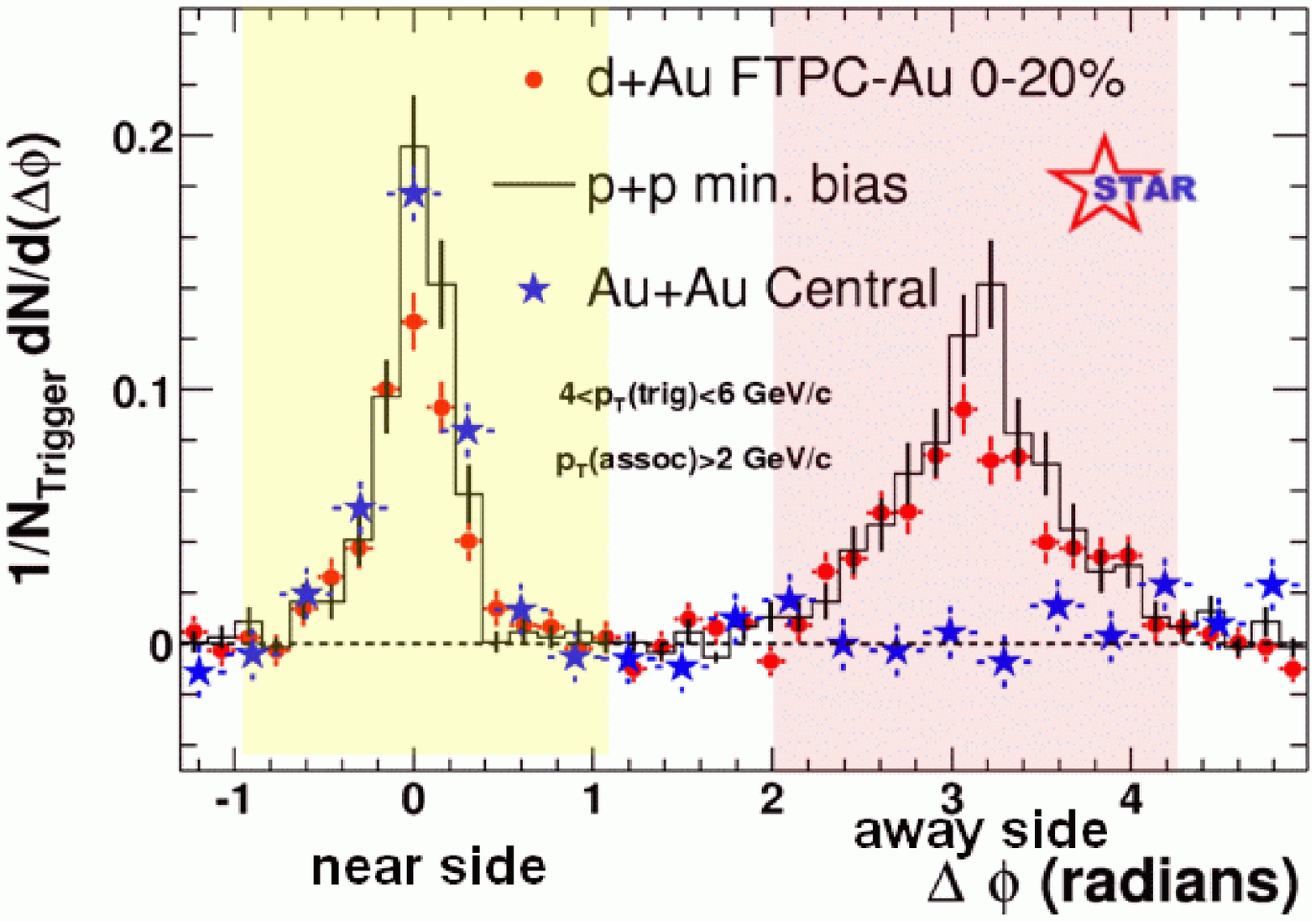}
\caption{Particles correlated with a high p$_T$ trigger particle showing 
the jet-like structure. For pp and dAu collisions, one can see the recoil
jet clearly. For central AuAu collisions - the recoil jet
disappears. \cite{awaysidestar}}
\label{fig:awaysidestar}
\end{minipage}
\end{figure}

In the third year of data taking, an important ``null'' experiment was
done in which deuteron-gold collisions were studied. It was important
to establish that the suppression of high p$_T$ particles was the result of
final state interactions which would be an
indication of the formation of a QGP and not due to some alteration
of the initial state such as a CGC.  Again, PHENIX looked at
mid-rapidity, central collisions. Fig~\ref{fig:raa} shows the
quantity R$_{AA}$ for $\pi^0$'s similar to R$_{CP}$ described above,
but using pp collision data in the denominator. One sees for central
Au-Au collisions, a factor of 4-5 suppression at high p$_T$, 
whereas for deuteron gold collisions the ratio is about unity. 
There is a slight indication of a Cronin type enhancement in the dAu 
collisions.

At this point, it was clear to RHIC experimentalists, that we were
observing a final state phenomena whose most probable explanation was
the loss of energy due the the passage of partons through a dense
partonic medium.  Calculations done by Vitev and
Gyulassy\cite{quench_models} reproduced the data reasonably well and
gave an energy density of 15 GeV/fm$^3$ early in the history of the
expanding fireball, with and initial gluon density of about
$\frac{dN_{gluon}}{dy}\sim 1100$.  Reasonable hadronic calculations
are unable to reproduce such a large energy loss\cite{hadronic}.

One can then further study the loss of energy by looking at opposite
side jets, since one of the hard partons would traverse a larger
distance in the dense medium. This was done by triggering on a high
p$_T$ particle and looking at the opposite side in the collision. In
peripheral collisions (or in pp collisions) the opposite side jet
signal should be rather strong, whereas in central Au-Au collisions,
the opposite side jet would be considerably broadened with a large
multiplicity of soft particles resulting from the energy loss of the
outgoing parton.  The typical p$_T$ of particles from collisions
assuming a thermal distribution would be well below 1.5 GeV.
Fig.\ref{fig:awaysidestar} shows a correlation plot in azimuthal
angle. An initial trigger particle was chosen with the requirement
that the p$_T$ be between 4 and 6 GeV. The angle between all particles
with p$_T$ above 2 GeV/c and the trigger particle are then
plotted. For pp and dAu collisions, one sees a clear two ``jet''
structure. In contrast, while the same side jet appears clearly in
central Au-Au collisions, the away side jet disappears.  The simplest
explanation is that the hard processes occurs near one surface. One of
the jets escapes with very little energy loss while the other jet is
almost completely quenched. The particles associated with the away
side jet can be identified if the correlation is extended down to
very low p$_T$ (0.15 GeV) since momentum must be conserved.  These
particles are very soft, higher in multiplicity and broader in angle;
in short they approach a thermalized distribution as one might
expect if the phenomenon is really due to the energy loss of partons
in a colored plasma.

\section{Hadronization}
All of the processes discussed so far are amongst quarks and gluons.
However, we see hadrons in our detectors and not quarks and gluons. The process
of hadronization is one of the most interesting aspects of the study
of QCD at RHIC, since this involves chiral symmetry breaking - or the 
generation of hadronic mass, and confinement.

\begin{figure}
\centering
\includegraphics[width=3in]{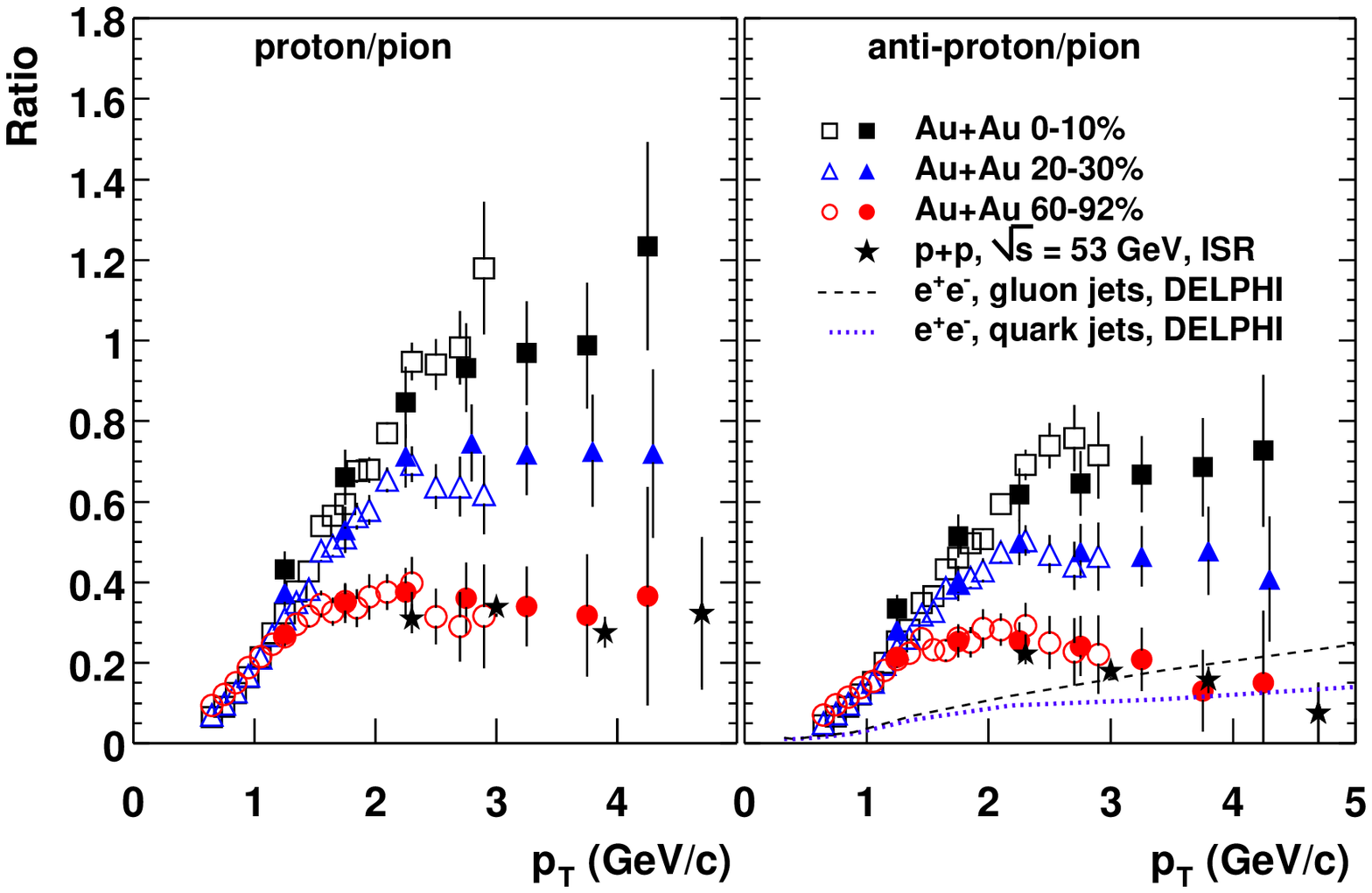}
\hspace{.0in}%
\includegraphics[width=1.9in]{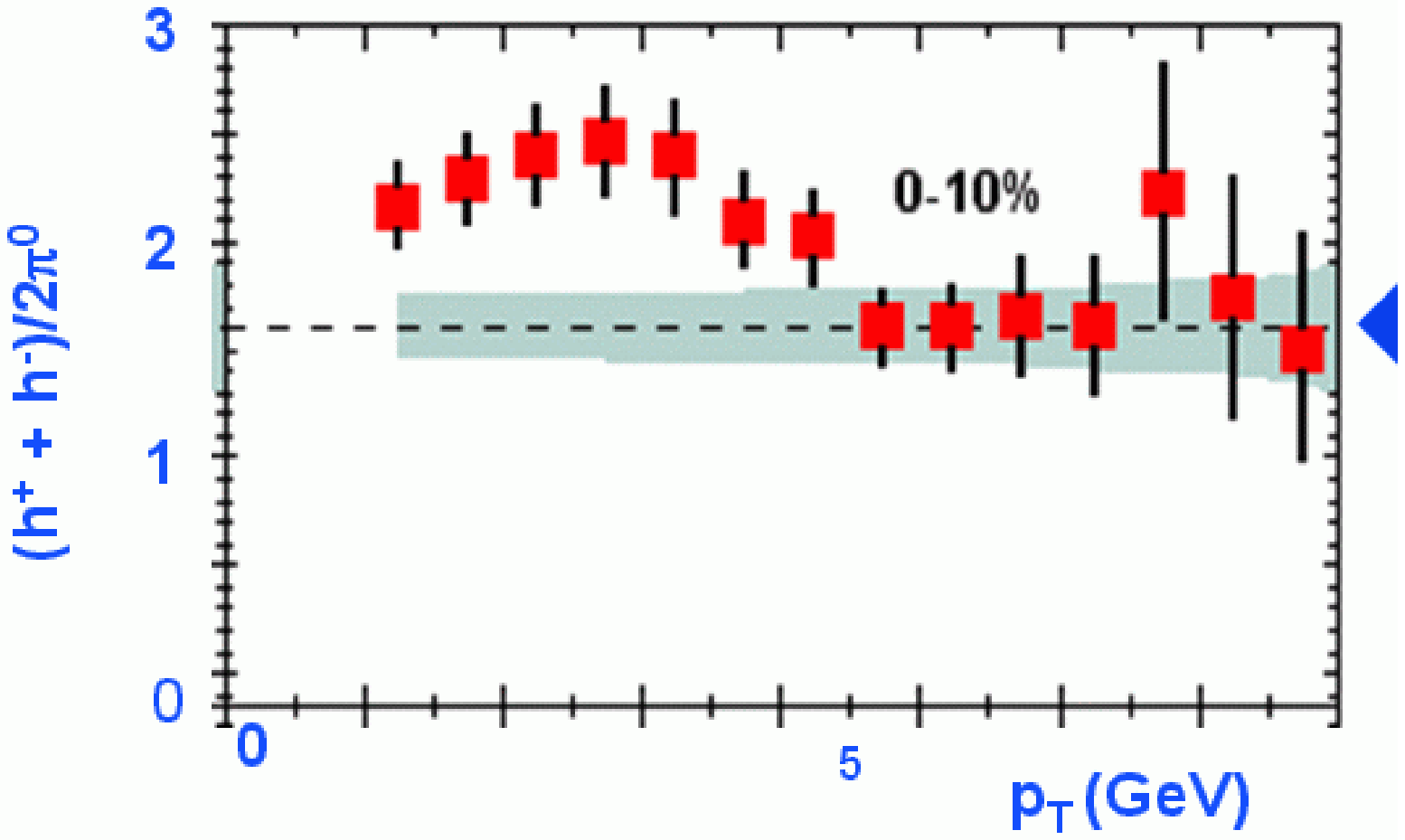}
\label{fig:htopi}
\caption{Left: (Anti)proton to $\pi^{0}$ ratio for different centrality classes for
Au+Au collisions at $\sqrt{s_{NN}}=200~$GeV.  Error bars represent the
quadratic sum of statistical and systematic errors (PHENIX).  Right: charged hadron
to 2$\pi^{0}$ ratio showing the return to the nominal value above 5 GeV/c (PHENIX).}
\end{figure}

One of the curious puzzles that faced the experiments was the large
proton to pion ratio at moderate p$_T$'s between 2 and 5 GeV.
Critical to this measurement was the particle identification
capabilities of the PHENIX experiment. Fig. \ref{fig:htopi} shows the
(anti)proton to pion ratio.  For central events the ratio is about 1.0
for protons, and 0.8 for anti-protons.  For peripheral events the
values are similar to that from pp collisions and
jets\cite{ptopi}. The data extends to about 4.5 GeV where PHENIX's
time of flight is no longer able to uniquely identify protons. In
order to check if this behavior extends to higher momentum, the
charged hadron to neutral pion ratio was measured. The charged hadron
is a mixture of charged pions which one can assume is about twice the
neutral pion yield, (anti)protons, and kaons.  Above 5 GeV this ratio
returns to a nominal value of about 1.5 consistent with pp collisions,
so the effect is confined to a p$_T$ range between 2 and 5 GeV. Such a
large production of baryons at moderate p$_T$ contradicts our current
understanding of fragmentation in the vacuum where only about 20\% of
the particles are baryons. This led theorists to assume there was some
mechanism for hadronization which depended on the density. They assumed
that hadrons were forming from a recombination of quarks already
present in the medium\cite{recomb}. Such a mechanism would enhance
baryons at high p$_T$. If one assumes there is a ball of thermal
partons expanding from internal pressure, the joining of 3 quarks
(baryons) would create a hadron with a p$_T$, 3/2 that of a hadron
created from 2 quarks (mesons). Since the spectrum is
falling steeply, this would lead to a large enhancement of baryons.

\begin{figure}
\begin{minipage}[t]{0.48\linewidth}
\centering
\includegraphics[width=2.5in]{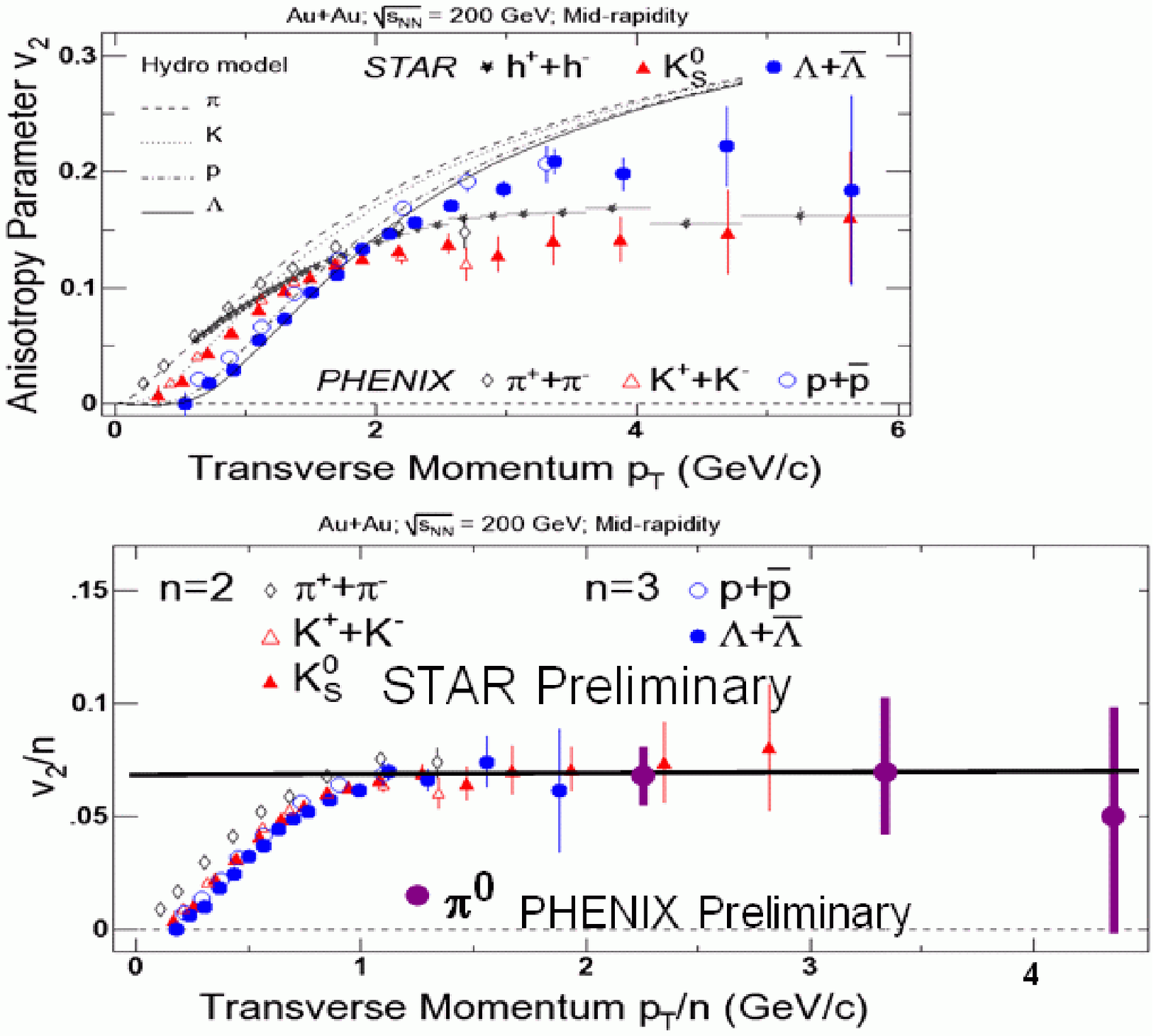}
\caption{Top: v$_2$ for various particles as a function of p$_T$.
Bottom: v$_2$ vs p$_T$ both scaled by n, the number of valence quarks, showing that above 1 GeV, all fall on the same line.}
\label{fig:v2scaled}
\end{minipage}%
\hspace{.2in}%
\begin{minipage}[t]{0.48\linewidth}
\centering
\includegraphics[width=2.4in]{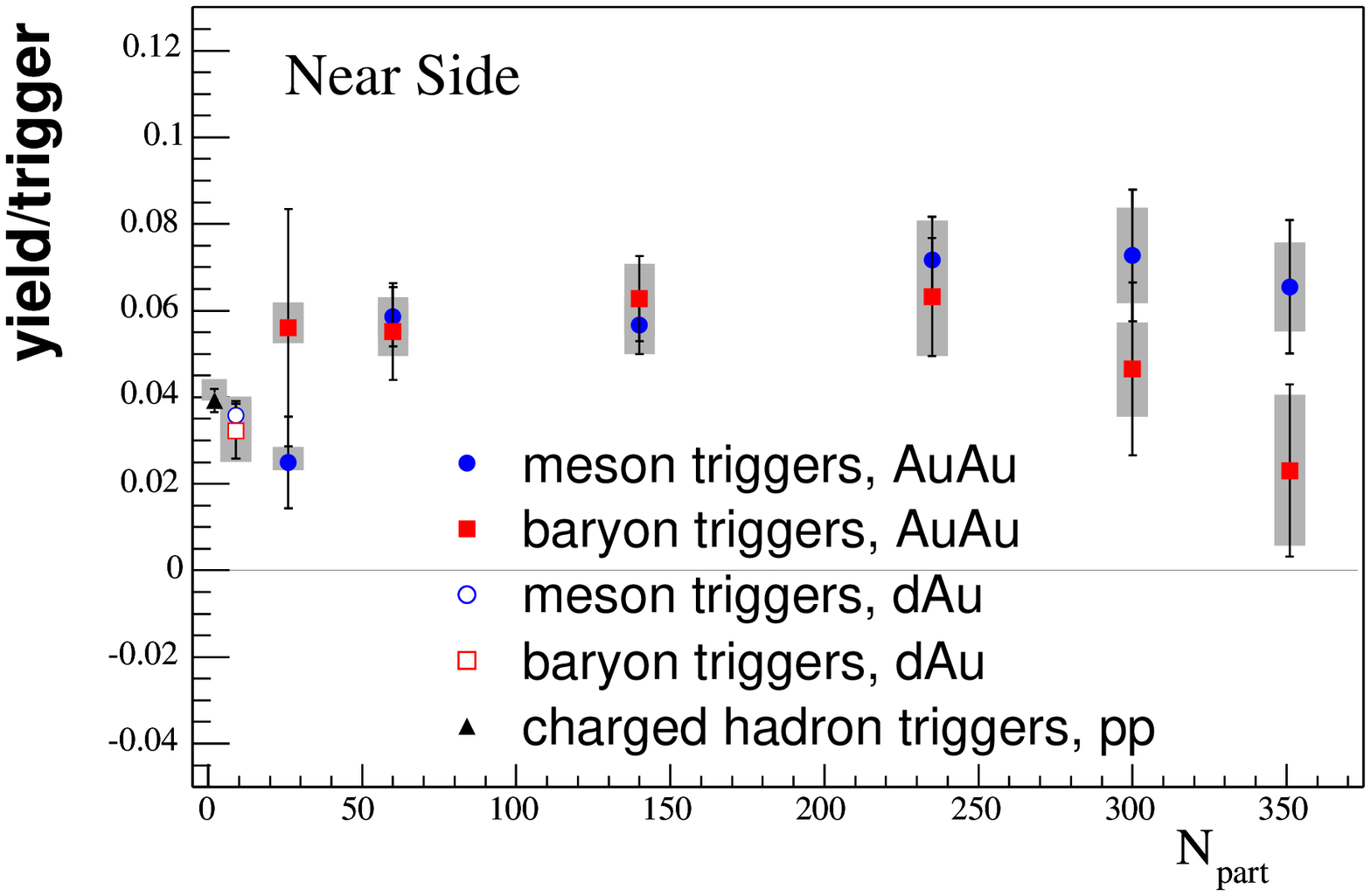}
\caption{Yield of particles withing a cone around a high p$_T$ trigger 
particle as a function of centrality as described in the text (PHENIX).}
\label{fig:jetcorr}
\end{minipage}
\end{figure}

One can test this idea a second way. Elliptic flow (v$_2$) develops
early in the collision when the degrees of freedom are presumably
quarks and gluons. If recombination is at work, then the elliptic flow
of identified particles, scaled by the number of constituent quarks
should reflect the underlying elliptic flow of partons.
Fig~\ref{fig:v2scaled} shows the v$_2$ of a variety of particles vs
p$_T$.  After rescaling with n, the number of valence quarks, all
hadrons fall on the same line above p$_T$/n $\sim$ 1 GeV.  In fact the
idea works down to very low p$_T$ for all particles aside from the
pions. One of the causes of this discrepancy is that many of the low
p$_T$ pions are actually from the subsequent decays of resonances such
as the $\Delta$ and $\rho$. More complex models which actually include
such effects bear this out\cite{ko}.

A question now arises. Is recombination from thermal quarks or do
fragmentation quarks from hard collisions also participate? If
recombination is from purely thermal quarks, then hadrons formed from
recombination should show no jet-like correlations.
Fig.~\ref{fig:jetcorr} shows the centrality dependence of the
associated charged hadron yield for particles between $1.7<p_T<2.5$
GeV/c above a combinatorial background for trigger baryons and trigger
mesons in the p$_T$ range 2.5-4.0 GeV/c in a $54^\circ$ cone around
the trigger particle\cite{jetcorr}.  Both mesons and baryons, which
presumably are made via recombination in the momentum range in
question, have associated particles, meaning that they have some
jet-like qualities to them. This appears to mean that by some
mechanism, hard scattered partons, hadronize by picking up partners
from the thermal bath. The radiated gluons (which are primarily
collinear with the fast quark) retain some of their
directionality. For baryons in central collisions, the effect appears
to be reduced, hinting that the formation of baryons in the most
central collisions is primarily from thermalized quarks.

The rather simple picture of recombination as a means of
hadronization, while appealing, leaves some open questions.  First,
the quark degrees of freedom have a place in this picture - but where
are the gluons? In this picture, it is as if the quarks already have
their dressed constituent masses before recombining. How did this
happen? Where is chiral symmetry broken? There is also a seeming
reduction of entropy inherent to the mechanism. Where does it go? By
its very nature this picture is a schematic cartoon for
non-perturbative physics. Mueller and his colleagues\cite{recomb}
among others have tried to put it on a more theoretically sound
footing, however this requires assumptions (e.g. ignoring the masses
of the hadrons) which limits the direct applicability of the theory to
p$_T > 2$ GeV or so.

\section{Conclusion}
The first three years data taking at RHIC have yielded a wealth of
information.  There is strong evidence that 1) the system has reached
a very high energy density, greater than 10 times that of an ordinary
nucleon 2) the system thermalizes rapidly 3) the system behaves as a
liquid of near zero viscosity, indicating that it is very strongly
interacting 4) the system is very opaque indicating that the cross
sections are extremely large - much larger than is typical of hadronic
cross sections. The simplest explanation of these phenomena is that
RHIC has formed is a thermalized system of quarks and gluons. It
may be that the degrees of freedom are more complicated. Indeed
recombination ideas may indicate that the degrees of freedom may
change as the system passes through the phase transition. 

Operationally, both experimentalists and theorists at RHIC no longer
think of the degrees of freedom as ordinary hadrons, but rather as a
near thermalized system of quarks and gluons- a Quark-Gluon Plasma.
The challenge now is that of characterizing the system that is
created.  We still have very little experimental understanding of the
chiral symmetry and deconfinement transitions.  One of the most direct
probes of the mass is the dilepton decay of light mesons ($\rho,
\omega$, and~$\phi$). These resonances have short lifetimes. The
invariant mass of electrons whose source is the decay of resonances
inside the high temperature fireball, should reflect the mass of the 
resonance in a high temperature vacuum. In
addition, we have only begun to probe the charm sector. A high
statistics run including the $J/\psi$ has been completed and analysis
is now beginning which may yield more insights into the mechanism of
confinement.  Direct photon - quark pairs can be used to as a
calibrated probe to make more quantitative measurements of energy
loss. Probes such as HBT correlations have yet to be understood.  All
of these probes are being pursued, and will be needed to understand
the fundamental connection between the QCD vacuum, the masses of the
hadrons, and the chiral and deconfinement phase transitions.

\end{document}